\documentclass[twocolumn,showpacs,amsmath,amssymb,prl]{revtex4}
\usepackage{graphicx}
\usepackage{dcolumn}
\usepackage{bm}

\begin{document}
\input{epsf}

\title{An Electromagnetic Signature of Supermassive Black Hole Binaries\\
That Enter Their Gravitational-Wave Induced Inspiral}

\author{Abraham Loeb}

\affiliation{Astronomy Department, Harvard University, 60 Garden St.,
Cambridge, MA 02138, USA}

\begin{abstract} 
Mergers of gas-rich galaxies lead to black hole binaries that
coalesce as a result of dynamical friction on the ambient gas. Once
the binary tightens to $\lesssim 10^3$ Schwarzschild radii, its merger
is driven by the emission of gravitational waves (GWs). We show that
this transition occurs generically at orbital periods of $\sim$1--10
years and an orbital velocity $v$ of a few thousand ${\rm km~s^{-1}}$,
with a very weak dependence on the supply rate of gas ($v\propto
\dot{M}^{1/8}$).  Therefore, as binaries enter their GW-dominated
inspiral, they inevitably induce large periodic shifts in the broad
emission lines of any associated quasar(s). The probability of finding
a binary in tighter configurations scales as $v^{-8}$ owing to their
much shorter lifetimes.  Narrow-band monitoring of the broad emission
lines of quasars on timescales of months to decades can set a lower
limit on the expected rate of GW sources for LISA.
\end{abstract}

\date{\today}
\maketitle

\paragraph*{Introduction.} Recent advances in general relativistic 
simulations of the final coalescence phase of black hole (BH) binaries
through gravitational wave (GW) emission \cite{Pre}, sparked new
interest in related electromagnetic (EM) signals \cite{Loeb}.  New
observational searches were initiated for wide BH binaries in their
pre-merger phase or single quasars which are displaced from their host
galaxy spatially or spectroscopically due to GW recoil after the
merger \cite{Com}.  In addition to testing the recent numerical
predictions, such searches can be used to calibrate the expected rate
of GW sources for the proposed LISA
mission \cite{lisa}, as well as to reduce the error
box on the source localization after LISA is launched. The next
generation of simulations will attempt to incorporate gas dynamics and
radiative transfer in order to calculate the EM counterparts to GW
sources \cite{Cent}. If these EM signatures are sufficiently unusual
relative to typical quasar variability, one could search for them well
before LISA is launched.  Since BH binaries spend more time at larger
separations, wider binaries should be more abundant in quasar
surveys. In this paper we study a generic EM signature of the
widest quasar binaries whose dynamics is dominated by GW emission.

The formation of BH binaries is a natural by product of galaxy mergers
(which are generic to the hierarchical build-up of galaxies in the
standard cosmological model), since almost all galaxies in the local
Universe are observed to harbor a BH at their center \cite{Gut}.  The
tidal torques generated during a merger extract angular momentum from
any associated cold gas and concentrate the gas near the center of the
merger remnant, where its accretion onto a BH results in a bright
quasar \cite{quasar}.

\paragraph*{Binary Parameters and Statistics.} 

We consider two black holes with masses $M_1$ and $M_2$ in a circular
orbit of radius $a$ about each other \cite{Pr}. Their respective
distances from the center of mass are $a_i=(\mu/M_i)a$ ($i=1,2$),
where $\mu=M_1M_2/M$ and $M=M_1+M_2$. We define the parameter
$\zeta=4\mu/M$, which equals unity if $M_1=M_2$ and is smaller
otherwise. The orbital period is given by,
\begin{equation}
P=2\pi (GM/a^3)^{-1/2}=1.72~{\rm yr}~a_{16}^{3/2} M_8^{-1/2},
\end{equation}
where, $a_{16}\equiv (a/10^{16}~{\rm cm})$ and $M_8 \equiv
(M/10^8M_\odot)$.  The angular momentum of the binary can be expressed
in terms of the absolute values of the velocities of its members $v_1$
and $v_2$ as $J=\Sigma_{i=1,2}M_iv_ia_i=\mu va$, where the relative
orbital speed is
\begin{equation}
v=v_1+v_2=(2\pi a/P)= 1.15\times 10^4~{\rm km~s^{-1}} M_8^{1/2}
a_{16}^{-1/2}~.
\end{equation}

We focus our discussion on mergers of gas-rich galaxies since those
are most likely to result in bright quasars which are easy to
observe. Hydrodynamic simulations of such mergers indicate that the
associated BH binaries inspiral steadily as a result of dynamical
friction on the ambient gas \cite{Escala}. In contrast, BH binaries in
gas-poor mergers, which shrink by scattering stars, may stall once
they reach a separation $a\lesssim 1$pc if the supply time of new
stars which cross their orbit exceeds the Hubble time (resulting in
the so-called ``final parsec problem'' \cite{review}).  In gas-rich
mergers, the rate of inspiral slows down as soon as the gas mass
interior to the binary orbit is smaller than $\mu$ and the enclosed
gas mass is no longer sufficient for carrying away the entire orbital
angular momentum of the binary, $J$ \cite{Bence}.  Subsequently,
momentum conservation requires that fresh gas will steadily flow
towards the binary orbit in order for it to shrink.  The binary
tightens by expelling gas out of a region twice as large as its orbit
(similarly to a ``blender'' opening a hollow gap \cite{Milos}) and by
torquing the surrounding disk through spiral arms
\cite{Hayasaki,Mac,Cuadra}.  Fresh gas re-enters the region of the
binary as a result of turbulent transport of angular momentum in the
surrounding disk. Since the expelled gas carries a specific angular
momentum of $\sim va$, the coalescence time of the binary is inversely
proportional to the supply rate of fresh gas into the binary region.
In a steady state, the mass supply rate of gas that extracts angular
momentum from the binary, $\dot{M}$, is proportional to the accretion
rate of the surrounding gas disk.  Given that a fraction of the mass
that enters the central gap accretes onto the BHs and fuels quasar
activity \cite{activ,Dot}, it is appropriate to express $\dot{M}$ in
Eddington units $\dot{\cal M}\equiv \dot{M}/\dot{M}_E$, corresponding
to the accretion rate required to power the limiting Eddington
luminosity with a radiative efficiency of $10\%$, $\dot{M}_E=2.3
M_\odot~{\rm yr}^{-1}M_8$.  We then find,
\begin{equation}
t_{\rm gas}\approx (J/ \dot{M} va)={\mu /\dot{M}}= 1.1\times 10^7~{\rm yr}~
\zeta \dot{\cal M}^{-1} .
\label{tga}
\end{equation}
For a steady $\dot{\cal M}$, the binary spends equal amounts of time
per log $a$ until GWs start to dominate its loss of angular
momentum.

The coalescence timescale due to GW emission is given by \cite{Shapiro},
\begin{equation}
t_{\rm GW} = {5\over 256} {c^5a^4\over G^3 M^2\mu}=2.53\times 10^5~{\rm yr}~
{a_{16}^4 \over \zeta M_8^3} .
\end{equation}
By setting $t_{\rm GW}=t_{\rm gas}$ we can solve for the orbital 
speed, period, and separation at which GWs take over,
\begin{eqnarray}
v_{\rm GW}&=& 7.2\times 10^3~{\rm km~s^{-1}}~\zeta^{-1/4} (\dot{\cal M}
M_8)^{1/8}~;\label{vgw}\\ 
P_{\rm GW}&=& 7.1~{\rm yr}~\zeta^{3/4} M_8^{5/8} \dot{\cal M}^{-3/8}~;\label{pgw}\\ 
a_{\rm GW}&=& 8.3\times 10^{-3}~{\rm pc}~\zeta^{1/2}M_8^{3/4}\dot{\cal M}^{-1/4}.\label{agw}
\end{eqnarray}
For a binary redshift $z$, the observed period is $(1+z)P_{\rm
GW}$. Remarkably, the orbital speed at which GWs take over is very
weakly dependent on the supply rate of gas, $v_{\rm GW}\propto
\dot{M}^{1/8}$. It generically corresponds to an orbital separation of
order $\sim 10^3$ Schwarzschild radii ($2GM/c^2$).  

In an equal-mass binary, each BH moves relative to the center-of-mass
at a speed of ${1\over 2} v_{\rm GW}$.  Coincidentally, this orbital
speed happens to be similar to the velocity width of the broad
emission lines (BEL) of quasars \cite{Peterson}. This fortunate
coincidence implies that the time-dependent gravitational potential
and UV illumination of the binary members will introduce variability
into the spectroscopic shape of any associated BEL.  It is believed
that BEL are emitted by numerous compact clouds which cover about a
tenth of the sky of the quasar \cite{Peterson}.  Even if the BEL
clouds originate in the outer edge of the accretion disk around one of
the BHs, their dynamics will be affected by the combined gravitational
force and UV illumination from both BHs.

We therefore robustly predict that {\it when BH binaries from mergers
of comparable-mass, gas-rich galaxies start to inspiral by GW
emission, they would show variability with a period of months to
decades (for $M(1+z)^{8/5}\sim 10^6$--$10^9M_\odot$) in any associated BEL,
across a range of Doppler velocities comparable to the BEL spectral
width}.

The notion that BH binaries would display periodic shifts in their
emission lines was recognized a long time ago \cite{Gaskell}.  So far,
intensive monitoring programs for a limited number of active galactic
nuclei revealed spectroscopic variability but no reproducible evidence
for a periodic signal \cite{Halpern}.  The new insight added by
Eqs. (\ref{vgw}-\ref{agw}) is that BH binaries should generically
start their GW-induced inspiral at orbital periods that are
observationally accessible, and at orbital velocities that are
comparable to the BEL width. This would naturally lead to a
double-peaked line profile which can be observed to vary
periodically. In contrast, the line profile is expected to remain
singly peaked \footnote{Disk emission could also lead to a double
peaked line profile for a single BH but with no orbital variability
(e.g., M. Eracleous, {\it et al.}, Astrophys. J. {\bf 490}, 216
(1997)). } at the lower orbital velocities associated with wider
binary separations (where the superposed BEL of the two BHs are only
slightly offset in velocity relative to each other), or for binaries
that are much more compact than their common BEL region (and hence act
gravitationally as a single BH).

The periodic Doppler shift of the BEL is expected to be smaller for
unequal mass binaries.  If each BH controls the spectroscopic centroid
of its own BEL feature, the BEL associated with $M_i$ will be shifted
by $v_i=(\mu/M_i) v$.
The relative abundance of BH binaries with different mass ratios can
be modeled theoretically \cite{vol}.

The probability of observing periodic BEL shifts due to a binary
transitioning to a GW-dominated inspiral, ${\cal P}_{\rm GW}$, depends
on the unknown fraction of the associated quasar lifetime which
overlaps with this final phase of binary coalescence. In case
accretion is suppressed by the opening of an inner cavity in the disk
\cite{Milos}, it is possible that no BEL emission is observable early
in the GW-dominated inspiral phase. However, recent numerical
simulations \cite{Hayasaki,Mac,Cuadra,activ,Dot} indicate that
accretion should continue during this phase through the formation of
two secondary accretion disks around the individual BHs.  If quasar
emission persists throughout the coalescence process at a constant
$\dot{\cal M}$ due to the abundant supply of cold galactic gas, then
one is likely to find as many binaries with separation $a_{\rm GW}\sim
10^{-2}~{\rm pc}$ as with a separation of tens of pc (for which the
enclosed gas mass is comparable to $\mu$). In that case, the prospects
for discovering a quasar binary with a periodic BEL shift are good,
since an example for a tight binary at a projected separation of $\sim
7.3~{\rm pc}$ has already been identified \cite{Rod} and three sub-pc
candidates have been reported recently \cite{Bog}.

The probability of finding binaries deeper in the GW-dominated regime,
${\cal P}\propto t_{\rm GW}$, diminishes rapidly at increasing orbital
speeds, with ${\cal P}={\cal P}_{\rm GW}(v/v_{\rm GW})^{-8}$.

\paragraph*{Discussion.}
State-of-the-art numerical simulations \cite{Cuadra,Dot} focused so
far on wide binary separations where gravitational torques dominate
the interaction of the binary with the surrounding gas, but did not
attempt to describe the late, viscosity-driven, evolutionary phase of
interest here.  This late inspiral phase is mediated by angular
momentum transport in the circumbinary disk which delivers fresh gas
from large distances into the binary region.  Since quasar accretion
disks on scales of $\sim 10^4$ Schwarzschild radii are
radiation-pressure dominated, a proper numerical treatment of the
turbulent transport of angular momentum requires a
radiation-magneto-hydrodynamics code. Suitable numerical simulations
were only applied to a single BH so far \cite{Turner}, but could be
extended with a global code \cite{glob} to the binary accretion
problem.

Given the limitations of existing simulations, one is restricted to
modeling the inflow of gas into the binary orbit through analytic
approximations.  A recent analytic treatment of viscous transport
within a circumbinary accretion disk \cite{Bence} ignored the
existence of individual compact accretion disks around the two BHs and
adopted a simplified prescription for $\alpha$-viscosity and binary
migration.  The formation of two compact accretion disks around the
BHs (that is necessary for producing BEL) is expected to occur through
gas inflow from the inner edge of the circumbinary disk \cite{Haya2},
and the moving inner disks could affect the gas flow around the binary
both hydrodynamically \cite{activ} and radiatively (especially if they
radiate near the Eddington limit \cite{Dot}).  Bearing these
simplifications in mind, Fig. 3 in Ref. \cite{Bence} is in excellent
agreement with Eq. (\ref{vgw}), confirming that the GW-driven phase is
approached at a characteristic orbital velocity of $\sim 10^4~{\rm
km~s^{-1}}$ with only a weak dependence on the accretion model
details.
The generic results of Eqs. (\ref{vgw}-\ref{agw}) naturally lead to
double-peaked line profiles which vary periodically on timescales of
months to decades.

Future monitoring of the broad emission lines of quasars can set a
lower limit on the expected rate of GW sources for LISA. A multi-year
search dedicated to periodic variability of BEL in a large sample of
quasars has not been performed as of yet.  Preliminary studies have
already made use of spectroscopic data from SDSS \cite{sdss} and
SDSS-III \cite{sdss3}.  Any related discovery would motivate the
implementation of narrow-band filters in photometric variability
surveys, culminating with the future LSST \cite{lsst}.  A precursor
for this approach may be realized in the near future.

The Palomar Transient Factory \cite{ptf} (PTF) plans to acquire
H$\alpha$ filters at 6564\AA~which are 80--100\AA~wide. The periodic
Doppler shift of the BEL in quasar binaries that enter their
GW-dominated inspiral corresponds to a fractional wavelength change of
$(v_{\rm GW}/c)\sim 1\%$ or $\sim 66$\AA. It would be possible to
notice such a shift as an observed line slides in and out of the above
mentioned filters.  The sky surface density of quasars brighter than
an i-band magnitude of 19, for which the Mg II, C IV or C III lines
are redshifted to 6564 \AA, is $\sim 0.2$ per square degree on the
sky. This results in 3 quasars per pointing if two filters are
used. The planned survey will monitor up to several hundred fields
containing up to $\sim 10^3$ quasars with a cadence of 1 month, and
measure relative i-band fluxes to a precision of 0.3\% (3
milli-magnitudes). It is also possible to design a dedicated survey
that will cover 40\% of the sky (a few thousand quasars) with a 1-year
cadence.  Separately from this program, a survey is underway to study
variability (reverberation mapping) of quasars brighter than an r-band
magnitude of 21 using six 200\AA~wide filters of the COMBO-17 survey
\cite{Wolf} on the Chandra Deep Field South, using the ESO 2.2 m
telescope at La Silla \cite{maoz}.

Spectroscopic surveys are much more labor intensive.  Previous
spectroscopic searches for BEL variability did not report any
detection of periodicity for several tens of quasars that were
monitored over many years \cite{Botti} or in a sample of a few
thousand quasars that were sampled only a few times over less than a
year \cite{Wil,vari}, but identified examples of longer-term
variability without a measured period \cite{Gaskell}.

Since quasar BEL are known to vary for other reasons \cite{Halpern},
it is crucial to identify periodicity before classifying a candidate
source as real. In confirmed cases, measurements of the period and
velocity shift can be used to solve for $M$ and $a$ up to an ambiguity
regarding the inclination angle of the orbital plane relative to the
line of sight. This ambiguity may be removed through reverberation
mapping \cite{Peterson,vari}, since a measurement of the time delays
between temporal variations in the UV continuum originating near the
BHs and the two response times of their BEL, would provide an
independent estimate of $M_1$ and $M_2$ (and hence $\zeta$). Even
before periodic variability is observed, reverberation studies can be
used to distinguish the double line profile of a binary (for which the
two lines would respond differently to variations in the UV continuum
of the two quasars \cite{Pet2}) from that of a single disk emitter
(for which the two lines will vary simultaneously).  In addition, the
quasar luminosity could be used to estimate $\dot{M}$ based on related
hydrodynamic simulations \cite{Hayasaki,Mac,Cuadra,activ,Dot}. With
all this information, observers will be in a position to examine
whether $t_{\rm gas}$ is indeed of order $t_{\rm GW}$ in these BH
binaries.

\paragraph*{Acknowledgments.}
I thank Y. Birnboim, L. Blecha, B. Kocsis, E. Ofek, and N. Stone for a
careful reading of the manuscript and helpful comments.  I also thank
the Joint Space science Institute at the Univ. of Maryland for
organizing a workshop that inspired this paper.  This work was
supported in part by NSF grant AST-0907890.

\end{document}